\newcommand{\diracslash}[1]{#1\llap{/\kern2pt}}

\newcommand{\be}{\begin{equation}}
\newcommand{\ee}{\end{equation}}
\newcommand{\bea}{\begin{eqnarray}}
\newcommand{\eea}{\end{eqnarray}}
\newcommand{\ba}[1]{\begin{array}{#1}}
\newcommand{\ea}{\end{array}}

\documentclass[superscriptaddress,secnumarabic,amssymb,nobibnotes,nofootinbib,aps,pre,showpacs]{revtex4}
\usepackage{graphicx}
\begin{document}
\title{On the Foundations of Quantum Mechanics: Wave-Particle Non-Duality and the Nature of Physical Reality}
\author{N. Gurappa}
\email{dr.n.gurappa@gmail.com }
\affiliation{Velammal College, Chennai, Tamil Nadu, India}

\begin{abstract}
The Schr\"odinger's wave function can naturally be realized as an `{\it instantaneous resonant spatial mode}'  in which quantum particle moves and hence the Born's rule is derived after identifying its origin. This realization facilitates the visualization of `{\it what's really going on?}' in the Young's double-slit experiment which is known to be the central mystery of quantum mechanics. Also, an actual mechanism underlying the `{\it spooky-action-at-a-distance}', another mystery regarding the entangled quantum particles, is revealed. Wheeler's delayed choice experiments, delayed choice quantum eraser experiment and delayed choice entanglement swapping experiments are unambiguously and naturally explained at a single quantum level without violating the causality. The reality of Nature represented by the quantum mechanical formalism is conceptually intuitive and is independent of {\it the measurement problem}.
\end{abstract}

\maketitle

Quantum mechanics is an extremely successful theoretical description of Nature, especially in the sub-atomic world where the classical mechanistic concepts seem to fail completely.  Nevertheless, for more than ninety years, there is no consensus about what kind of physical reality is being revealed by the quantum formalism irrespective of its ability to predict accurately the exact outcomes of various experiments. According to Prof. Feynman, the central mystery of quantum mechanics is contained in the Young's double-slit experiment which is about the wave-particle duality of a single quantum \cite{Feyn1}. Twenty years later, he once again declared that the entanglement of two or more particles is one more deep mystery in the quantum world \cite{Feyn2}. It is not only important but also unavoidably necessary to conceptually visualize the true picture of reality described by the quantum formalism not only for solving the above mentioned mysteries, but also for further progress in fundamental physics like quantum gravity, unification of fundamental forces, quantum cosmology e.t.c.

In Young's double-slit experiment, a monochromatic source emits coherent light which passes through a double-slit assembly to a detector screen where an interference pattern reminiscent of wave nature is formed. On the other hand, photoelectric effect, Compton effect, Raman effect, e.t.c., strongly suggests the existence of particle nature of light. The usual intuition about particle is that it is a localized entity present at some definite position in space, whereas the wave is a delocalized one and hence they are incompatible with each other. But, light seems to possess both natures simultaneously; however, only one nature seems to be observable at a given moment. These mutually exclusive natures of light's behavior is generally known as wave-particle duality. Not only light, but all material particles like electrons, protons, atoms, molecules {\it e.t.c.}, are known to exhibit the wave-particle duality \cite{spi1,spi2,spi3,spi4,spi6}. The quantum formalism itself never imposes any limitations on its validity only to microscopic objects and it can, in principle, be applied to materials of any scale.

\begin{figure}[htbp]
\centerline{\includegraphics[width=0.8\textwidth]{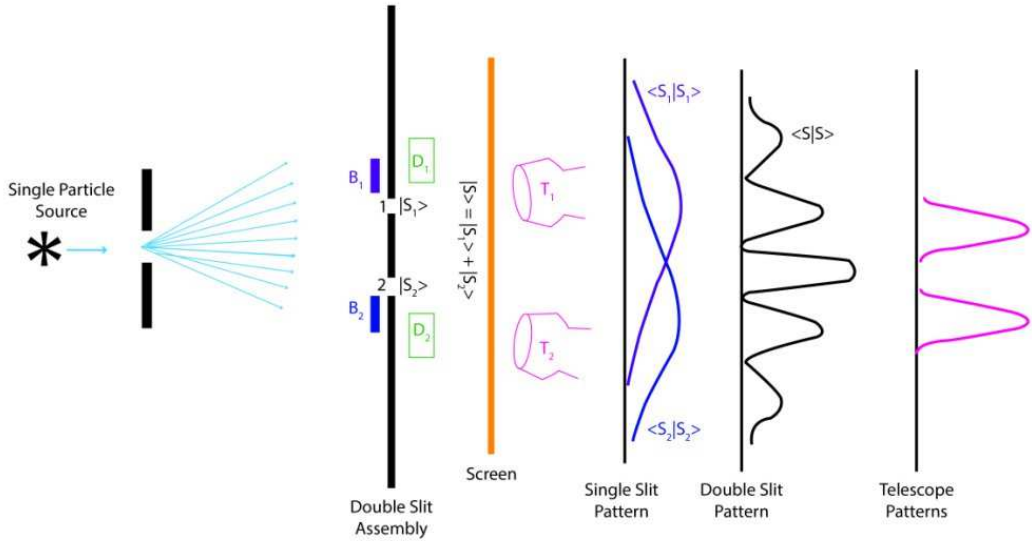}}
\caption{\textbf{Diagram of single-particle Young's double-slit
experiment}:~A source shoots single particles, one at a time,
towards a double-slit assembly. $1$ and $2$ represents two slits
through which the state vectors $|S_1>$ and $|S_2>$ emerge out and
get superposed as $|S> = |S_1> +|S_2>$. $B_1$ and $B_2$ are two
blockers which can block either slit $1$ or $2$ at any time. $D_1$
and $D_2$ are two detectors useful to find out through which slit
any particle is passing to the screen. Immediately behind the
screen, two telescopes, $T_1$ and $T_2$, are placed such that
particles passing through $1$ and $2$ reach $T_1$ and $T_2$,
respectively. After collecting a large number of individual
particles, the observed particle distribution pattern on the screen
and telescopes were also shown in the same diagram. When slit $2$
($1$) is blocked, a distribution $<S_1|S_1>$ ($<S_2|S_2>$) is
obtained. If both the slits are open, then the observed distribution
is $<S|S>$.} \label{Figure1}
\end{figure}

See the Fig.~\ref{Figure1} below representing the Youngs's double-slit experiment.
Consider the single particles to be photons. In this case, each photon is fired at the double-slit one-at-a-time such that the time interval between consecutively fired photons may be greater than the time of arrival of any one photon from the source to the screen. This assures that, each and every photon is really independent and they don't know each other. As a large number of photons are being collected on the screen, an interference pattern gradually emerges. If slit-1 (slit-2) is blocked, then a clump pattern corresponding to single slit diffraction of slit-2 (slit-1), supposed to be of particle nature, appears on the screen. This suggests that every individual photon is aware of whether  one or both slits are open. The interference pattern suggests to infer that a single particle-like photon `somehow' passes through both the slits simultaneously. A surprise occurs when a detector observes through which slit a photon is really passing through. It always appears as going through slit-$1$ or slit-$2$ like a particle but never through both the slits simultaneously like a wave. However, now the interference pattern disappears and two clump patterns appear which look like a proof for the observed particle behavior at the respective slits. This is generally known as quantum enigma i.e., when photons are watched, they appear to go through only one slit like particles. But, when they are not watched, then they seem to go through both the slits simultaneously like a wave.

Now, consider the Wheeler's delayed-choice situation \cite{Wheeler}. The screen is removed quickly exposing the twin telescopes, after a photon has already passed  through the double slits. The interference pattern which would have occurred on the screen is lost and two clump patterns, one at each telescope, are formed. Using wave-particle duality, the inference drawn is that the photon retroactively rearranges its past history of passing through both slits simultaneously like a wave to that of passing through only one slit at a time like a particle. Here, I would like to emphasized that the wave-particle duality implies retrocausality.

It's worth mentioning that a single quantum event is always observed as a particle-like well localized chunk in the detector. It is necessary to collect a large number of single events in order to build the interference pattern on the screen and hence the wave nature of a single quantum can be inferred \cite{spi1,spi2,spi3,spi4,spi6}. In other words, wave-particle duality demands a mechanism by which the wave passing through the double-slit becomes a particle on the screen. If the wave represents merely the probability of finding a particle at some location \cite{Born}, then such a mechanism is unnecessary; however, it leads to a non-intuitive conclusion that {\it the quantum particle is a `probability wave'} until it is observed as a particle. Otherwise, the interference pattern cannot occur on the screen. On the other hand, if the wave is considered to be physically real, then it's a must to have a definite mechanism for this wave to become a particle upon observation. Obviously, to provide such a mechanism, the Schr\"odinger equation, the very basic equation of quantum mechanics has to be modified \cite{GRW}.

There are various interpretations of quantum formalism, like, the Copenhegan interpretation \cite{Au}, the Bhomian mechanics \cite{Bohm}, many-worlds interpretation \cite{Au} etc, but none of them provides intuitively pleasing picture of reality. The Bohmian mechanics provides a beautiful picture of quantum phenomenon by removing wave-particle duality but, only at a cost of introducing epistemology-ontology duality, i.e., the wave function provides an epistemological probability distribution for the lack of knowledge about the initial position of a particle and at the same time, it plays an ontological role as a guiding wave for the same particle's motion. How some physical quantity can, at the same time, play a dual role as both epistemological and ontological is questionable. Further, these interpretations neither give a derivation nor a justification for the Born's rule which is intimately connected to our experimental experience and also to the measurement problem.

Another mystery of quantum mechanics, the entanglement, is an experimentally established physical phenomenon of Nature \cite{SAD1,SAD2,SAD3,SAD4,SAD5,SAD6,SAD7,SAD8}. It's well-known that two distinct quantum systems, separated by a long distance after some initial interaction is described by a joint entangled state such that any measurement on one system has an instantaneous influence on the outcome of another measurement made on the second system, which appears to be violating the cosmic speed limit imposed by the theory of relativity. Einstein called this instantaneous effect as `spooky action at a distance'  which demands a natural mechanism, within quantum formalism, for its existence.

Interpretation must provide a relation between the mathematics of the formalism and the physical world. The picture of quantum reality, i.e., wave-particle non-duality, presented here is not the one coming from the direct experience of the usual world but can be felt intuitively within the quantum mechanical formalism itself. In the present paper, only the case of time-independent non-relativistic quantum mechanics is considered because, its interpretation naturally goes through time-dependent and relativistic cases.

In the Newtonian paradigm, a point is located in 3D space by specifying its coordinates by a rigid measuring rod and hence attaching the property of rigidity to the space itself. Such a rigid but empty space provides an absolutely passive and unchanging `stage' for all physical phenomena happening. But in Einstein's general theory of relativity, space (actually space-time) is dynamical and can bend, stretch, twist and even ripple. It's active and dictates particles' movement to lie along the geodesic in the curved space-time. {\it It's very important identifying the actual space in which a physical phenomenon is happening. Otherwise, the reality start appearing to be strange, weird and counter-intuitive.}

Starting from the classic Stern-Gerlack experiment, all quantum phenomena appear to take place in a `complex' space rather than the usual 3D Euclidean space. Then why the  macroscopic objects, which are obviously composites of `quantum entities', appear to live in Euclidean space? So, it's clear that the space around us is indeed `complex' in nature even though it is `effectively' observed as Euclidean. Quantum mechanics is reveling a profound and remarkable property of space (or space-time) itself which is quite different from that of general relativity.

In the following, I list a few assumptions and a hypothesis needed to describe the nature of space in which quantum phenomena takes place. After that, Young's double-slit, Wheeler's delayed-choice, delayed-choice quantum erasure and delayed-choice quantum entanglement swapping experiments are explained within the quantum mechanical formalism at an individual quantum level.

\begin{enumerate}

\item {\underline{{\bf Empty Space:}}}
\begin{enumerate}
\item  {\it `Empty space' is in general an infinite dimensional complex vector space of continuous (and also discrete) dimensions}. Let ${\bf S}$ be the set of elements denoting the empty space. The most crucial property of ${\bf S}$ is
$$ {\bf S} = {\bf S} \otimes {\bf S} \quad ; {\bf S} \in {\bf S} $$
where, $\otimes$ stands for the direct-product and $\in$, for `is an element of'.

\item Any quantum mechanical Hilbert space, {\bf H}, is always a subset i.e., ${\bf H} \subset {\bf S}$.
\end{enumerate}

\item \underline{{\bf Hypothesis:}}

{\it A precise set of elementary particles in this infinite dimensional complex vector pace, {\bf S}, with well-defined properties and interactions among them results in the macroscopic
manifestation of matter with respect to which the eigenvalues of position operator effectively form the $3 D$ Euclidean space.}

\item \underline{{\bf Instantaneous Resonant Spatial Mode (IRSM):}}
\begin{enumerate}

\item When a free particle of definite momentum eigenvalue `$p$' is created, then a resonant spatial mode `$< {\bf r}|S>$' appears instantaneously in the entire space such that the particle's motion is completely confined within this mode, where $|S> \in {\bf H}$. This situation is exactly like the case in general relativity where a particle moves in a curved space-time. Since $|{\bf r}>$ and $< {\bf r}|S>$ are in one-to-one correspondence, and $|S>$ represents both the particle and its IRSM together, here onwards I call, without loss of generality, the state vector itself as an IRSM.

The IRSM can be visualized as follows:
\begin{eqnarray}
|S> = \int d {\bf r} |{\bf r}><{\bf r}|S>
\end{eqnarray}
At every eigenvalue ${\bf r}$ of the position operator $\hat{\bf r}$, attach the complex vector $|{\bf r}><{\bf r}|S> $. That's the visualization of IRSM in which, at some eigenvalue ${\bf {r_p}}$ the particle is present, carrying the corresponding vector  $|{\bf {r_p}}><{\bf {r_p}}|S> $. Since the vectors associated with the particle and any other spatial point
are identical in structure and nature, all these vectors at every position eigenvalue, ${\bf r}$, behaves exactly like the particle itself. That's why the term `resonance' and also because both the particle and its IRSM always appear or disappear simultaneously together, i.e., the IRSM survives as long as the particle survives with the same momentum `$p$'. Hence, the IRSM itself, even in the absence of particle,  can induce empty, but ontological empty modes for all possible physical processes provided those corresponding physical systems lie in the region of IRSM (As an example, see towards the end of the paper the delayed choice quantum erasure experiment for a single quantum particle).

\item  Now onwards, the quantum mechanical formalism is adopted as it is without any additions and modifications. In fact, the existence of quantum mechanical commutation relations like $[\hat{x} \,\,,\,\, \hat{p}] = i \hbar \,;$ where, $\hat{x}$ and $\hat{p}$ being position and momentum operators respectively, are the actual reason for the IRSM to appear along with a particle. The Plank's constant $h$ can be realized as a resonant coupling parameter between a particle and its IRSM. Here, coupling means that the particle is actually free to move but always confined within its IRSM, which is like a deterministic constraint. I am omitting related mathematical argument, because the present paper is aimed at an interpretation level. However, the de Broglie relation (or the dispersion relation in general) is an easiest way to look at it.

\item The initial boundary condition for IRSM is the point in the complex space where the momentum was originated and remains unaltered as long as the particle sustains with the same momentum, i.e., `origin remains unaltered as long as the particle gets unaffected' - is the important property of the complex vector space. The final boundary condition depends on where the particle will end up and need not be a fixed boundary condition (both boundary conditions are fixed for the case of bound systems). If the particle undergoes some momentum changing interaction, then the earlier IRSM disappears completely and a new IRSM corresponding to new momentum appears with origin at the spatial point where the particle gained new momentum. {\it This particular picture of a particle flying in its own IRSM is non-dualistic in nature, further irreducible and is independent of any measurement procedure}.

\item In the case when the particle is subjected to some potential, then the IRSM is a solution of the time-independent Schr\"odinger's wave equation with that potential and its behavior is exactly like a classical wave since it's a solution of a wave equation. Here, the energy eigenvalue changing interaction corresponds to the momentum changing interaction mentioned above.
\end{enumerate}

\item \underline{{\bf Inner-product as interaction:}}
\begin{enumerate}

\item This non-dualistic picture of a particle flying in its own IRSM is not analogous to any classical wave, though the IRSM obeys a wave equation. It's well-known that the
square of amplitude of a classical wave is proportional to its intensity. Therefore, such an intensity can't be claimed for the IRSM.

\item If the particle is going to end up, say for example on a detector screen, then a dual vector, $<S|$, is induced in that screen which interacts with the IRSM. The
interaction is given by the inner-product, $<S|S>$. Then, that particle will be observed at some location only in this region of inner-product. Note that, this inner-product interaction happens the moment the particle appears at the source. Instead of the detector screen, some
surface of a distant planet or an eye of some creature etc., can also be considered.

\item Precisely due to this inner-product interaction, we effectively feel the space to be $R^3$ Euclidean.

\item If the detector states do not have complete basis to span $|S>$ and is associated with a projection operator $\hat{P}$, then the IRSM seen by the detector is $|S_D> = \hat{P} |S>$ and the dual vector excited in it is $<S_D|$. Therefore, the interaction region for the particle detection is $<S_D|S_D>$.

\end{enumerate}

\item \underline{{\bf Principle of minimum phase and quantum jumps:}}

\begin{enumerate}

\item The state $|S>$ and $e^{i \phi} |S>$ will have different sense of direction in the complex vector space; where, $\phi$ is an absolute initial phase of $|S>$ at its origin and $|e^{i \phi}| = 1$.

\item Let us consider a classical scenario of tossing a coin. It is possible in principle to predict exactly whether head or tail occurs on a flat ground. If one is ignorant about some parameters involved in the dynamics of the coin, then probability can be invoked. Let a normal vector, $\hat{\bf n}$, passes through the center of the coin from tail side to head surface and $\theta$ be the angle between $\hat{\bf n}$ and any parallel axis to the ground surface. Just before landing, consider the position of the coin at a height $h \le r$ above the ground; where $r$ is the radius of the coin. If $\theta$ lies between $ 0 \le \theta \le \pi $, then head will be the out come. If $ - \pi \ge \theta \ge \pi $, then tail occurs i.e., depending upon in which range $\theta$ lies, the coin is forced to jump into either head or tail state. By replacing $\hat{\bf n}$ by electron's spin magnetic axis and the parallel vector to the ground by magnetic force direction in the Stern-Gerlac (SG) apparatus, then the resulting situation of the coin is exactly identical to that of an electron considered by Bell \cite{Bell}. Here, it should be noted that both the coin and electron were considered to be in $3D$ Euclidean space. But in reality, electron flies in its own IRSM. Let $|S>$ be the IRSM corresponding to the original spin state of electron in the complex vector space (spatial dependence is suppressed for simplicity). The unit operator along the direction of force in the SG apparatus is $1_{\rm op} = |\uparrow><\uparrow| + |\downarrow><\downarrow|$. Then the IRSM in the SG apparatus is
$$|S> = |\uparrow><\uparrow|S> + |\downarrow><\downarrow|S>$$
As one can see easily, akin to the coin in Euclidean space jumping into either head or tail depending upon $\theta$, the electron spin in complex vector space jumps into either $|\uparrow>$ or $|\downarrow>$ depending upon which of the complex numbers $<\uparrow|S>$ and $<\downarrow|S>$ has a minimum phase. It should be noted that even though the electron jumps into, say $|\uparrow>$, the empty mode $|\downarrow>$ still survives until the detection of the electron. Further, it's emphasized that, only the magnitude of the coefficients, $|<\uparrow|S>|$ and $|<\downarrow|S>|$, of the IRSM depend upon the angle between electron's spin and magnetic force direction while their phases determine to which state the electron actually jumps.

\item The IRSM can be decomposed into various components but not the particle moving in it. In other words, only the IRSM can undergo superposition but not the particle flying in it.

\item Through experiment, one observes only preexisting properties of the particles. However, the values of the observed properties may get altered due to the act of observation.

\item \underline{{\bf Derivation of Born's rule:}} When the IRSM, $|S>$, is decompose into eigenstates, $|a_i>$; $i = 1, 2, 3, \cdots$, of an operator $\hat{A}$, then the particle jumps from $|S >$ to one of the eigenstate, say $|a_p>$, such that the `phase' of the complex number $<a_p|S >$ is minimum when compared with all other phases made by the remaining eigenstates. Note that, all other empty eigenstates are still ontologically present though the particle itself is in the minimum phase eigenstate, $|a_p>$. If detection takes place, then the particle will be found in $|a_p>$ with an eigenvalue $a_p$ because all other empty modes do not contribute. This indeed coincides with the `projection postulate' or `the reduction of state vector'. The IRSM is
\begin{eqnarray}
|S> = \sum_i |a_i><a_i|S>
\end{eqnarray}
and it interacts with the excited dual-mode, $<S|$, in the detector as
\begin{eqnarray}
<S|S> = \sum_i <S|a_i><a_i|S> = \sum_i |<a_i|S>|^2 \rightarrow |<a_p|S>|^2
\end{eqnarray}

As it's well-known, the eigenvalues $a_p$ are the direct measurable but not $|<a_p|S>|^2 $ which can be computed by repeating the same experiment several times with the same kind of particles. Each particle is represented by the same $|S>$ but with a different initial phases i.e., $e^{i \phi} |S>$. So, when the IRSM is decomposed into eigenstates of $\hat{A}$, then each particle sees a minimum phase with some particular eigenstate $|a_i>$ depending upon $\phi$ i.e., the phase of the complex number $e^{i \phi} <a_i|S>$. Each value of $\phi$ gives raise to a different minimum phase with different eigenstate such that out of total number of particles with random phases, a fraction of them will be found in $|a_i>$. In the limit of total number of particles tending to infinity, one can see that the fraction coincides with $|<a_i|S>|^2 $,
\begin{eqnarray}
<S|S> = \sum_i <S|a_i><a_i|S> = \sum_i |<a_i|S>|^2 =  1
\end{eqnarray}
which is the Born's rule.

\item Here, it's worth mentioning a quote by Dirac \cite{Dirac}, "{\it Question about what decides whether the photon is to go through or not and how it changes its direction of polarization when it goes through can not be investigated by experiment and should be regarded as outside the domain of science}". The present non-dualistic picture completely agrees with the first part of the statement, "...can not be investigated by experiment" - because $<S|S>$ will not give any experimentally detectable phase information, but disagrees with the last part, " ...and should be regarded as outside the domain of science".

Therefore,
\begin{eqnarray}
{\rm Frequency \,\,of \,\, observation} = {\rm Probability\,\, in\,\, {\rm Quantum} \,\, {\rm Mechanics}}  \nonumber
\end{eqnarray}
Therefore, the conclusion is that quantum mechanics itself is not a probabilistic theory. Probability or the frequency of occurrence arises due to the nature of doing experiment.
A remark is that the probability interpretation actually leads to the measurement problem and is the actual obstacle preventing the visualization of physical reality. Once one attaches probability to some physical process and describes it in terms of probability, then such a physical process can not be explained independent of observation or measurement device. In the present non-dualistic picture, there is no probability in Nature. Nevertheless, it is not possible to predict the future event in the effective 3D Euclidean space which we perceive, because the actual quantum phenomenon is taking place in the complex space and the absolute phase of the IRSM is unavailable to experimental observation due to the inner-product interaction. So, we are forced to observe the frequency of outcomes after an infinite number of repeated, identical measurements. The true nature of our existing space, i.e., complex  can be inferred by analyzing the mutually incompatible experimental outcomes.

\item Consider three sequential detectors A, B and C represented by observables $\hat{A}$, $\hat{B}$ and $\hat{C}$ \cite{Sakurai}. Let $|a_i>$, $|b_j>$ and $|c_k>$ be the  eigen vectors of the operators $\hat{A}$, $\hat{B}$ and $\hat{C}$ with eigenvalues $a_i$, $b_j$ and $c_k$, respectively; where $i, j, k = 1, 2, 3, \cdots$. Let the detectors $A$, $B$ and $C$ select some particular $|a^\prime_i>$, $b^\prime_j>$ and $|c^\prime_j>$ and rejects the rest. Let $|S>$ be the IRSM in which the quantum particle is flying. The first detector A has its own vectors space spanned by the eigenstates $|a_i>$. When $|S>$ is subjected to A, it gets resolved into various components as
\begin{eqnarray}
|S> = \sum_i <a_i|S> |a_i>
\end{eqnarray}
in which only one component $|a^\prime_i>$ is allowed to come out and the rest are blocked. If the initial phase of $|S>$ is such that it makes a minimum angle with $|a^\prime_i>$, then the particle will be present in this mode and passes on to the next detector B.  When the mode $|{{\tilde{a}}_i}>$ $ (\equiv <a^\prime_i|S> |a^\prime_i>)$ encounters the detector B, then it gets resolved in B's space as $|{\tilde{a}_i}> = \sum_j <b_j|{{\tilde{a}}_i}> |b_j>$. Now, B allows only $|b^\prime_j>$ by blocking the rest, hence $ <b^\prime_j|{\tilde{a}}_i> |b^\prime_j>$ will encounter C. This vector in C space is
\begin{eqnarray}
<b^\prime_j|{{\tilde{a}}_i}> |b^\prime_j> = <b^\prime_j|{{\tilde{a}_i}}> \sum_k <c_k|b^\prime_j> |c_k> \,\,.
\end{eqnarray}
Now the detector C allows only $|c^\prime_k>$ to come out. There should be another detector D which measures the outcome from C. So, in response to the mode $<b^\prime_j|{{\tilde{a}_i}}> <c^\prime_k|b^\prime_j> |c^\prime_k>$, a dual mode $<{{\tilde{a}_i}}|b^\prime_j> <b^\prime_j|c^\prime_k> <c^\prime_k|$ is excited in D and interacts as given by the inner-product, i.e.,  $|<b^\prime_j|{\tilde{a}}_i>|^2 |<c^\prime_k|b^\prime_j>|^2$. If the initial phase of $|S>$ is such that the particle in it passes through all detectors A, B and C, then it will be detected at D. If one sends a large number of particles, all are described by the same $|S>$ but with different initial phases, then out of total number of particles, the fraction detected by D is given by
\begin{eqnarray}
<{\tilde{a}}_i|{\tilde{a}}_i> \rightarrow   |<b^\prime_j|{{\tilde{a}}_i}>|^2 |<c^\prime_k|b^\prime_j>|^2 \label{1R}
\end{eqnarray}

 Suppose, the detector B allows all its modes. Then C will encounter the superposition of these modes, i.e., $\sum_j <b_j|{\tilde{a}}_i> |b_j> = |{\tilde{a}}_i>$. In B space, though the particle is present in a particular mode $|b_j>$, all other empty modes do present ontologically and if unblocked, they will contribute at C. So, we have $|{\tilde{a}}_i> = \sum_k <c_k|{\tilde{a}}_i> |c_k>$. When C allows only one component $|c^\prime_k>$ to pass through, then D encounters a mode, $<c^\prime_k|{\tilde{a}}_i> |c^\prime_k>$. Then the excited dual mode $<{{\tilde{a}_i}}|c^\prime_k> <c^\prime_k|$ in D interacts as $|<{\tilde{a}}_i|c^\prime_k>|^2$. So, the frequency of observation at detector D is
\begin{eqnarray}
<{\tilde{a}}_i|{\tilde{a}}_i> \rightarrow |<{\tilde{a}}_i|c^\prime_k>|^2
\end{eqnarray}
which is very different from Eq. (\ref{1R}). Therefore, ontological presence of an empty mode has a physically observable effects.

\end{enumerate}
\item \underline{{\bf `No quantum jump' - situation:}}

When the state $|S>$ representing the IRSM of a particle is decomposed into various orthogonal eigenstates of an operator with continuous eigenvalues, then the particle, without any quantum jump, will naturally enter into one of the eigenstate whose phase is exactly same as that of $|S>$.

As an example, consider the space spanned by the eigenvalues, ${\bf r}$, of the position operator $\hat{{\bf r}}$ with eigenstates $|{\bf r}> $; where, ${\bf r} = \{x,y,z\}$.
$$|S> = \int d{\bf r} |{\bf r}> <{\bf r}|S>$$
Then the particle will be present in a state $|{\bf {r_p}}> <{\bf {r_p}}|S>$ whose phase is exactly same as $|S>$. Therefore, upon observation,
\begin{eqnarray}
<S|S> = \int d{\bf r} <S|{\bf r}> <{\bf r}|S> \rightarrow |<{\bf {r_p}}|S>|^2
\end{eqnarray}
Even though the particle is present in $|{\bf {r_p}}> <{\bf {r_p}}|S>$ with an eigen value ${\bf r}_p$, it's easy to see that when it hits the screen, the excited dual mode is $<S|$ but not
$<S|{\bf {r_p}}> <{\bf {r_p}}|$ because the detector is not projecting out this particular state, $|{\bf {r_p}}>$ . Therefore, the interaction is given by $|<{\bf {r_p}}|S>|^2$. (Note that, here one can treat the state vector either in Schr\'odinger picture or the position operator).

\item \underline{{\bf Outer-product:}}

\begin{enumerate}
\item All vectors in the Hilbert space, represented in the position basis, are super-imposed on top of each other and can co-exist in the same region of Euclidean space spanned
by the eigenvalues of the position operator.

Consider the eigenvalue equation for the position operator, $\hat{\bf r}$,
$$\hat{\bf r} |{\bf r}> = {\bf r} |{\bf r}>$$
Note that, the position eigenvalues ${\bf r}$ span the $R^3$ Euclidean space. But, there are higher order operators with tensor product states as eigen vectors with same eigen value ${\bf r}$, i.e., for example,
$$ (\hat{\bf r} \otimes I_{op} + I_{op} \otimes \hat{\bf r}) (|{\bf r}> \otimes |{\bf r}>) = {\bf r} (|{\bf r}> \otimes |{\bf r}>) $$
where, $I_{op} = \int d {\bf r} |{\bf r}><{\bf r}|$, is the unit operator in the position eigen space. Here, $(|{\bf r}> \otimes |{\bf r}>)$ is an example of superimposed state at ${\bf r}$. In other words, in a given region, any number of spatial modes can coexist and this corresponds to the outer-product of the state vectors. Classical waves, like ripples on a surface of water, do not behave this way and also this situation is in contrast
with the classical Newtonian and Einsteinan spaces which are uniquely described by the position eigenvalue ${\bf r}$ alone.

\item If a single particle is described by a tensor product state, then the form of this vector is independent of its direction in the complex vector space.

\item Consider an entangled state of two particles
\begin{eqnarray}
|S_{12}>> \equiv |S_1> |S_2> \label{ES}
\end{eqnarray}
obeying some definite conservation law
\begin{eqnarray}
(\hat{S}_1 + \hat{S}_2) |S_{12}>> = 0 \label{CL}
\end{eqnarray}
 If the state $|S_1>$ appears to be decomposed into two components during a physical process, then in reality it's not $|S_1>$ which splits into components, but the state $|S_{12}>>$, i.e., if
$$|S_1> \rightarrow |S_{1,a}> + |S_{1,b}> $$
then
\begin{eqnarray}
|S_{12}>> = |S_{1,a}> |S_{2,a}> + |S_{1,b}> |S_{2,b}> \label{ESDC}
\end{eqnarray}
such that the Eq. (\ref{CL}) holds for each component independently; where, $a$ and $b$ in the subscripts stand for two corresponding components. This splitting happens even
in the case of an empty mode.

\item \underline{{\bf Spooky action at a distance:}}

The description given here is essentially the EPR argument \cite{EPR} rewritten in the present non-dualistic interpretation. One can write the entangled state in Eq. (\ref{ES}) as
\begin{eqnarray}
|S_{12}>> &=&  \int_0^\infty d {\bf r}  |{\bf r}><{\bf r}|S_1> \otimes \int_0^\infty d {\bf r}  |{\bf r}><{\bf r}|S_2> \nonumber\\
 &=& \left(\int_0^\infty d {\bf r} \otimes  \int_0^\infty d {\bf r} \right)  \left\{(|{\bf r}><{\bf r}|S_1>) \otimes (|{\bf r}><{\bf r}|S_2>) \right\}
 \end{eqnarray}
At at every eigenvalue ${\bf r}$, the super-imposed states $(|{\bf r}><{\bf r}|S_1>) \otimes (|{\bf r}><{\bf r}|S_2>)$ couple to each other such that the Eq. (\ref{CL})
is strictly obeyed. Thus any change in the state $|S_1>$ at ${\bf r}$ instantly affects $|S_2>$ at the same ${\bf r}$ and hence everywhere. Also, in order to establish the conservation law in the  Eq. (\ref{CL}), the particles should have interacted at some common region which is taken to be the zero eigenvalue of the position operator. That's why the lower limit for both integrals above are  zero. This initial condition implies a constraint for any two arbitrary position operators acting on $|S_{12}>>$ as given by
$$(\hat{\bf {r_1}} \otimes I_{op} - I_{op} \otimes \hat{\bf {r_2}}) = 0$$
If ${\bf r_{p_1}}$ and ${\bf r_{p_2}}$ are locations of first and second particles, then observation on any one particle or on both particles simultaneously results in
\begin{eqnarray}
<<S_{12}|S_{12}>> \rightarrow |<{\bf r_{p_1}}|S_1>|^2 . |<{\bf r_{p_2}}|S_2>|^2
\end{eqnarray}
where, $<<S_{12}|$ is the excited dual mode in the detector.
So, the original IRSM $|S_{12}>>$ completely disappears leaving the particles in the correlated states subjected to the conservation law at every value of ${\bf r}$ and position operator constraint between any two values of ${\bf r_1}$ and ${\bf r_2}$, respectively. Suppose, only the first particle was observed but not the second one. Then, whatever the observed position eigenvalue of the first particle, ${\bf {r_{p_1}}}$, the second particle acquires the correlated eigenvalue ${\bf {r_{p_2}}}$ which becomes the origin for a new IRSM of the particle 2.

\item Let Alice makes a measurement on particle 1 in the discrete basis $|P_1^{A +}>$ and $|P_1^{A -}>$ and Bob makes a measurement on on particle 2 in his such a basis, $|P_2^{B +}>$ and $|P_2^{B -}>$. Therefore, one can write,
\begin{eqnarray}
|S_{12}>> =  <P_1^{A +}|S_{12}>> |P_1^{A +}>  +  <P_1^{A -}|S_{12}>> |P_1^{A -}>
\end{eqnarray}
and also
\begin{eqnarray}
|S_{12}>> =  <P_2^{B +}|S_{12}>> |P_2^{B +}>  +  <P_2^{B -}|S_{12}>> |P_2^{B -}>
\end{eqnarray}

If Alice finds the particle 1 in $|P_1^{A +}>$, then the resultant state gained by particle 2 is $<P_1^{A +}|S_{12}>> = |P_2^{A_1 +}>$, as mentioned above in 6.(d). But Alice would have found the particle in $|P_1^{A -}>$ as well. Then the state of particle 2 is $<P_1^{A -}|S_{12}>> = |P_2^{A_1 -}>$. This can be written in a single equation as
\begin{eqnarray}
|P_2^{A_1 \pm}> <P_1^{A \pm}|S_{12}>> \equiv |P_2^{A_1 \pm}>
\end{eqnarray}

Now, Bob will measure $|P_2^{A_1 \pm}>$ in his basis as
\begin{eqnarray}
|P_2^{A_1 \pm}> =  <P_2^{B +}|P_2^{A_1 \pm}> |P_2^{B +}>  +  <P_2^{B -}|P_2^{A_1 \pm}> |P_2^{B -}>
\end{eqnarray}
then correlation is
\begin{eqnarray}
C_2^{{B_2} \pm}({A_1} \pm) =  |<P_2^{B \pm}|P_2^{{A_1} \pm}>|^2
\end{eqnarray}

If on the same entangled pair, Bob would have measured the particle 2 first, then due to the instantaneous nature of the resonant mode, the outcome of Alice would have been
\begin{eqnarray}
C_1^{{A_1} \pm}(B_2 \pm) =  |<P_2^{A \pm}|P_2^{{B_2} \pm}>|^2
\end{eqnarray}
though, both these measurements exclude each other.

Suppose, Alice and Bob filters all directions of measurement except one, say, $|P_1^{A +}>$ and $|P_2^{B +}>$, then only in this case one can have
\begin{eqnarray}
C_2^{{B_2} +}({A_1} +)  = C_1^{{A_1} +}(B_2 +)
\end{eqnarray}
which is as well valid for an ensemble of identical entangled pairs. Suppressing the plus sign and particle indices,
\begin{eqnarray}
C^{{B} }({A} )  = C^{{A} }(B) = |<P^A|P^B>|^2 \equiv C(A,B) \label{Malus}
\end{eqnarray}
This is nothing but the Malus law and is the heart for the correctness of violation of Bell's inequality \cite{Bell}. As one can see from the above derivation, either local or non-local
hidden variables related to classical probability distributions can not reproduced this result.

\item The above derivation goes through even when the particles are prepared in a definite initial state, say, $|S_{12}^i>>$. Now, the generic two particle state $|S_{12}>>$ can be written as
\begin{eqnarray}
|S_{12}>> = |S_{12}^i>> <<S_{12}^i|S_{12}>> + |{\bar{S}}_{12}^i>> <<{\bar{S}}_{12}^i|S_{12}>>
\end{eqnarray}
where, $<<{\bar{S}}_{12}^i|S_{12}^i>> = 0$. Since the particles are prepared in $|S_{12}^i>>$,  by filtering the mode $|{\bar{S}}_{12}^i>>$, all measurements which detect the particle states will see
\begin{eqnarray}
|S_{12}>> \rightarrow |S_{12}^i>> <<S_{12}^i|S_{12}>>
\end{eqnarray}
So in the derivation given in 7.(e), instead of $|S_{12}>>$ one can use $|S_{12}^i>>$, whose overall factor $ <<S_{12}^i|S_{12}>>$ will not affect the final result in Eq. (\ref{Malus}). So, the conclusion drawn here is that the entangled state can be represented as a direct product state $|S_{12}>> = |S_1>|S_2>$ subjected to the conservation law $(\hat{S}_1 + \hat{S}_2) |S_{12}>> = 0$.

\end{enumerate}
\end{enumerate}

\section{Young's Double-Slit Experiment: What's really happening?}

Consider the Young's double-slit experiment as mentioned in Fig.~\ref{Figure1} . Here, the case of a single-photon shot onto the screen only after the registration of the previous photon is considered in order to elucidate the actual behavior of an individual quantum particle.

According to our classical intuition, if photons travel one by one, some through slit $1$ and some through slit $2$, then the naive expectation of them is to leave a pattern of two strips on the screen because they are thought to be flying in the passive $R^3$ space. Nevertheless, an interference pattern reminiscent of wave nature appears because the photons are not actually moving in the Euclidean space but in its own IRSM which obeys Schr\"odinger equation. As it is easy to see that, in the present non-dualistic interpretation,
macroscopic objects naturally yield clump patterns matching our classical intuition because, their de Broglie wave length is extremely small when compared to the size of the object, the dimensions of slits and their separation.

Let $|S_0>$ be the state vector representing the IRSM of a photon which started at the source. The projector, $\hat{P}_{\rm ds}$, associated with the double-slit assembly is
\begin{eqnarray}
\hat{P}_{\rm ds} = \int{d {\bf r}^{(1)}_1 |{\bf r}^{(1)}_1><{\bf r}^{(1)}_1|} + \int{d {\bf r}^{(2)}_2 |{\bf r}^{(2)}_2><{\bf r}^{(2)}_2|} \,\,,
\end{eqnarray}
where, $\{|{\bf r}^{(1)}_1>\}$ and $\{|{\bf r}^{(2)}_2>\}$ are position basis for slit $1$ and slit $2$, respectively. The state vector from double-slit to the screen is
\begin{eqnarray}
|S> = \hat{P}_{\rm ds}|S_0> = \int{d {\bf r}^{(1)}_1 |{\bf r}^{(1)}_1><{\bf r}^{(1)}_1|S_0>} + \int{d {\bf r}^{(2)}_2 |{\bf r}^{(2)}_2><{\bf r}^{(2)}_2|S_0>} \equiv|S_1> + |S_2>
\end{eqnarray}

The dual-mode excited in the detector screen is $<S|$ which interacts with the IRSM according to the inner product $<S|S>$.
\begin{eqnarray}
<S|S> = <S_1|S_1> + <S_2|S_2> +  <S_1|S_2> + <S_2|S_1>
\end{eqnarray}
Note that, the moment a photon appears at the source, its IRSM, through the double-slit, has already formed the interaction on the screen but remains unobservable until the photon's hit at some position, i.e., one photon contributes to one point in the interaction region, $<S|S>$.
The photon which is flying within the IRSM passes either through slit $1$ or $2$, depending on its initial phase. The moment photon's momentum changes either due absorption or scattering at the detector screen, then the entire IRSM disappears. This resembles the wave function collapse advocated in the Copenhegan interpretation but, without prescribing any  mechanism. Once the solution of Schr\"odinfer's wave equation is recognized as an IRSM, then the reduction of state-vector naturally exists within the quantum formalism itself.

The next photon appears at the source along with its IRSM whose phase will be different from the previous photon, i.e., its state may be given by $e^{i \phi} |S>$. The interaction region, $<S|S>$, is same for all photons but their hits on the screen occur randomly at different locations due to different values of phases, $\phi$. The randomness in phases is due to
its dependence on the detailed properties of the source and many other parameters, eventually like the particle's initial location in the entire Cosmos. After a large collection of photons landing at random positions, an interference pattern results on the screen which is nothing but the construction of the function $|<{\bf r}|S>|^2$ with individual points.

Let, the sets $\{|{\bf r}_1>\}$ and $\{|{\bf r}_2>\}$ be the position basis representing the slit $1$ and $2$ on the screen, respectively. Using unit operators
$\int{d {\bf r}_1} |{\bf r}_1><{\bf r}_1|$ and $\int{d {\bf r}_2} |{\bf r}_2><{\bf r}_2|$, one has,
\begin{eqnarray}
<S|S> &=& \int{d {\bf r}_1} <S_1|{\bf r}_1><{\bf r}_1|S_1> + \int d {\bf r}_2 <S_2|{\bf r}_2><{\bf r}_2|S_2> \nonumber\\
& & + \int \int d {\bf r}_1 {d {\bf r}_2} <S_1|{\bf r}_1><{\bf r}_1|{\bf r}_2><{\bf r}_2|S_2>  \nonumber\\
& & +\int \int {d {\bf r}_2} {d {\bf r}_1} <S_2|{\bf r}_2><{\bf r}_2|{\bf r}_1<{\bf r}_1|S_1> \nonumber\\
&=& <S_1|S_1> + <S_2|S_2> \nonumber\\
& & + \int \int {d {\bf r}_1} {d {\bf r}_2} <S_1|{\bf r}_1><{\bf r}_2|S_2> \delta({\bf r}_1 - {\bf r}_2) \nonumber\\
& & +\int \int {d {\bf r}_2} {d {\bf r}_1} <S_2|{\bf r}_2><{\bf r}_1|S_1> \delta({\bf r}_1 - {\bf r}_2)
\end{eqnarray}
If the slit bases are indistinguishable on the detector screen, {\it i.e.}, $ {\bf r}_1 = {\bf r}_2$, then the $\delta$-function contributes resulting in the interference. In the case the  bases are distinct, {\it i.e.}, $ {\bf r}_1 \ne {\bf r}_2$, then $\delta({\bf r}_1 - {\bf r}_2) = 0$ and the interference does not occur.

If the screen is replaced by the twin telescopes while a photon is in mid-flight, then
\begin{eqnarray}
(\hat{T}_1 + \hat{T}_2) |S> &=& (\hat{T}_1 + \hat{T}_2) (|S_1> + |S_2>) =  \hat{T}_1 |S_1> + \hat{T}_2 |S_2> \nonumber \\
&=& |\tilde{S}_1> + |\tilde{S}_2> \equiv |\tilde{S}>
\end{eqnarray}
where, $\hat{T}_1$ and $\hat{T}_2$ are operators associated with the telescopes, $T_1$ and $T_2$, since they are lens systems. Now, the old IRSM, $|S>$ is replaced to new IRSM $|\tilde{S}>$ but their origins remain unchanged i.e., they are same for both $|S>$ and $|\tilde{S}>$. From whatever be the position of photon during the replacement of $|S>$ to $|\tilde{S}>$,
 it continues to fly from there in the new IRSM, $|\tilde{S}>$. The photon's motion is always continues even though the IRSM itself changes suddenly. Later, I will use this important property of IRSM to explain the delayed choice entanglement swapping experiments. The continuity in photon's motion is governed by its conserved properties. Therefore, the observed photon distribution at telescopes is given by
\begin{eqnarray}
<\tilde{S}|\tilde{S}> = <\tilde{S}_1|\tilde{S}_1> + <\tilde{S}_2|\tilde{S}_2>
\end{eqnarray}
which is a clump-pattern.

Note that, {\it causality is preserved}, i.e., there is no retro-casual influences. {\it This underlying non-dualistic picture where `a particle always moves in its own IRSM' is independent of any measurement procedure}.

Let a detector $D$, using a state vector $|D>$, probe the photon through which slit it actually passes through. Now, one has,
\begin{eqnarray}
|S> \rightarrow |S^\prime>|D^\prime> + |S>|D> = \sum_{i=1}^2 ( |S^\prime_i> |D_i^\prime> + |S_i> |D> ) \equiv |S>>
\end{eqnarray}
where, $|S^\prime>|D^\prime>$ is an entangled state of the photon with the probe state and $<D|D^\prime_i> = 0$ and the resultant interaction on the screen is
\begin{eqnarray}
<<S|S>> &=& <S^\prime_1|S^\prime_1> <D_1^\prime|D_1^\prime> + <S^\prime_2|S^\prime_2> <D_2^\prime|D_2^\prime> + <S^\prime_1|S^\prime_2> <D_1^\prime|D_2^\prime> \nonumber\\
&+& <S^\prime_1|S^\prime_2> <D_2^\prime|D_1^\prime> \nonumber\\
&+& <S_1|S_1> + <S_2|S_2> + <S_1|S_2>  + <S_1|S_2>
\end{eqnarray}
If the photon interacts with the probe, then it enters into the state $|S^\prime> |D^\prime>$; hence, the photon will hit the screen at some point in the following reduced region of interaction
\begin{eqnarray}
<<S|S>> \longrightarrow && <S^\prime_1|S^\prime_1> <D_1^\prime|D_1^\prime> + <S^\prime_2|S^\prime_2> <D_2^\prime|D_2^\prime> \nonumber\\
 &+& <S^\prime_1|S^\prime_2> <D_1^\prime|D_2^\prime> + <S^\prime_1|S^\prime_2> <D_2^\prime|D_1^\prime>
\end{eqnarray}
Further, if $|D_1^\prime>$ and $|D_2^\prime>$ has a finite overlap with respect to the dual-vector space of the detector, then an interference patter occurs on the screen.

On the other hand, if the overlap is negligible, i.e., $<D_1^\prime|D_2^\prime> \simeq 0$, then the interference is lost and a clump pattern occurs as it can be easily seen from the above equation. This particular situation can be visualized without considering the entanglement with detector's probe as follows: Any momentum changing interaction of the photon with the detector's probe will result in the disappearance of the earlier IRSM, $|S>$, which had two origins at slit 1 and 2 at the same time. A new IRSM (either $|S_1^\prime>$ or $|S_2^\prime>$) corresponding to new momentum appears whose single origin lies at the interaction point in the space and it will not pass through any slit in the direction of photon's motion and interacts with the detector screen. The interaction is either $<S_1^\prime|S_1^\prime>$ or $<S_2^\prime|S_2^\prime>$. So, in the presence of detector, clump patters occur and when the detectors are taken away, the interference comes back. This particular property viz, `{\it the disappearance of interference pattern and the appearance of clump patterns whenever the photons are watched through which slit they are going}', is an ultimate proof for the underlying particle nature of  photons (or any other material particles). Nature is providing a double-confirmation for the particle nature, i.e., first, when it was observed through which slit it was passing and the second, then the disappearance of the interference pattern. If the underlying particle nature is absent, then disappearance of interference pattern is impossible.

In the case when the photon didn't interact with the probe, then it will be present in $|S>|D>$ and the observed pattern on the screen is
\begin{eqnarray}
<<S|S>> \rightarrow <S_1|S_1> + <S_2|S_2> + <S_1|S_2>  + <S_1|S_2>
\end{eqnarray}
Now, consider the case of the above Young's Double-Slit Experiment with Polarization filters $P_1$ and $P_2$ in the place of blockers $B_1$ and $B_2$ respectively. Then, the IRSM is given by
\begin{eqnarray}
|S>> = |S_1> |P_1> + |S_2> |P_2>
\end{eqnarray}
and the interaction of IRSM on the screen is
\begin{eqnarray}
<<S|S>>  = &&<S_1|S_1> <P_1|P_1> + <S_2|S_2> <P_2|P_2> \nonumber\\
&+&  <S_1|S_2> <P_1|P_2> + <S_2|S_1> <P_2|P_1>
\end{eqnarray}

Therefore, if $|P_1>$ and $|P_2>$ are orthogonal, then the interference vanishes.

Let $|P_1> = |H>$ and $|P_2> = |V>$. Then $|S>> = |S_1> |H> + |S_2> |V>$. Let us introduce a $45^0$ polarization rotator just before the screen. Then one has
\begin{eqnarray}
|S>> \rightarrow |\bar{S}>> = |{\bar{S}}_1> |\bar{H}> + |{\bar{S}}_2> |\bar{V}>
\end{eqnarray}
where, $|\bar{H}> = \frac{1}{\sqrt{2}} (|H> + |V>)$ and  $|\bar{V}> = \frac{1}{\sqrt{2}} (- |H> + |V>)$ , then on the screen one gets
\begin{eqnarray}
<<\bar{S}|\bar{S}>> &=& \frac{1}{2}  (<{\bar{S}}_1|{\bar{S}}_1> + <{\bar{S}}_2|{\bar{S}}_2> - <{\bar{S}}_1|{\bar{S}}_2> - <{\bar{S}}_1|{\bar{S}}_2>) <H|H> \nonumber \\
&+& \frac{1}{2} ( <{\bar{S}}_1|{\bar{S}}_1> + <{\bar{S}}_2|{\bar{S}}_2> + <{\bar{S}}_1|{\bar{S}}_2>  + <{\bar{S}}_1|\bar{S}]_2>) <V|V> \nonumber \\
&=&  <{\bar{S}}_1|{\bar{S}}_1> + <{\bar{S}}_2|{\bar{S}}_2>
\end{eqnarray}
Suppose, a Wollaston prism with unit operator $1_{\rm op} = |H><H| + |V><V|$ is introduced in between polarization rotator and screen, then it will resolve $|\bar{S}>>$ into two orthogonal components $|H>$ and $|V>$, which can be detected by two independent detectors $D_H$ and $D_V$; obviously the outputs are given by
\begin{eqnarray}
{D_H}_{\rm output} = \frac{1}{2}  (<{\bar{S}}_1|{\bar{S}}_1> + <{\bar{S}}_2|{\bar{S}}_2> - <{\bar{S}}_1|{\bar{S}}_2> - <{\bar{S}}_1|{\bar{S}}_2>)
\end{eqnarray}
and
\begin{eqnarray}
{D_V}_{\rm output} = \frac{1}{2} ( <{\bar{S}}_1|{\bar{S}}_1> + <{\bar{S}}_2|{\bar{S}}_2> + <{\bar{S}}_1|{\bar{S}}_2>  + <{\bar{S}}_1|{\bar{S}}_2>)
\end{eqnarray}
Note that, the polarization rotator can be introduced or removed randomly before a photon actually passes through it. As it was explained above in the Wheeler's delayed choice experiment, the photon continues to fly even during the IRSMs are being changed randomly. This method was used in an experiment by Jacques et al. \cite{Jacques}, which uses Mach-Zehnder interferometer instead of Young's double-slit assembly. Similar experiment was also done using single atoms \cite{SingleAtom}.

\section{Dopfer's experiment}
A source emits an entangled pair of particles with total zero momentum \cite{Zeilinger}. Call them as left particle and right particle such that their momentum entangled IRSM,
$|S>> = |L> |R>$, obeys
\begin{eqnarray}
({\hat{P}}_L + {\hat{P}}_R) |S>> =  0
\end{eqnarray}
where, $|L>$ and $|R>$ are IRSMs of left and right particles and ${\hat{P}}_L$ and ${\hat{P}}_R$ are their momentum operators respectively. This situation is identical to the one considered in Eq. (\ref{ES}).

\begin{figure}[htbp]
\centerline{\includegraphics[width=0.8\textwidth]{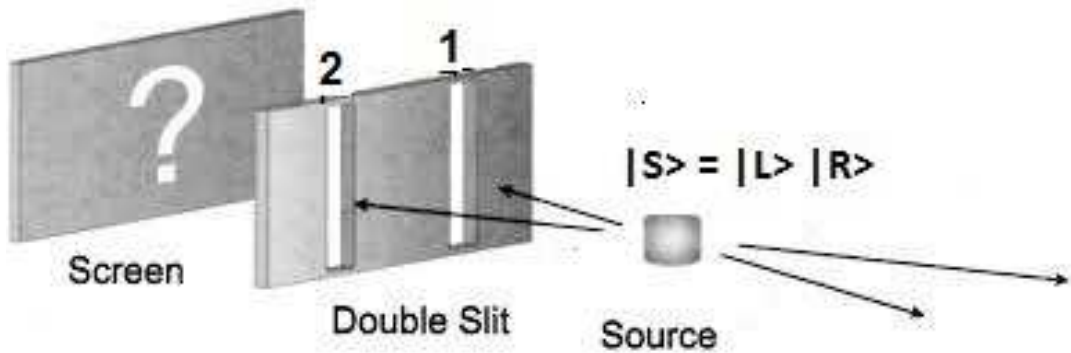}}
\caption{\textbf{Principle involved in Dopfer's experiment}:~The
moment a momentum-entangled particles are created at the source,
their IRSM, $|S>>$ $(= |L>|R>)$ gets excited in the entire space and
also through the double-slit. The IRSM, $|S^\prime>>$, through the
double-slit interacts with its dual $<<S^\prime|$ in the screen and
the left particle hits the screen in the region of interaction
$<<S^\prime|S^\prime>>$. Whether all such left particles exhibits
interference or clump pattern can be decided by an appropriate
measurement on the right particles.} \label{Figure3}
\end{figure}

The entangled IRSM $|S>>$ passes through the double-slit and reaches the screen as $|S^\prime>>$,
\begin{eqnarray}
|S^\prime>> = |L_1> |R_1> + |L_2> |R_2>
\end{eqnarray}

Note that by invoking the following mapping,
\begin{eqnarray}
|L_1> \rightarrow |a>_1, |R_1> \rightarrow |b>_2, |L_2> \rightarrow |a^\prime>_1, |R_2> \rightarrow |b^\prime>_2 \,\,{\rm and}\,\, |S^\prime>> \rightarrow |\psi> \nonumber
\end{eqnarray}
one gets
\begin{eqnarray}
|\psi> = |a>_1 |b>_2 + |a^\prime>_1 |b^\prime>_2
\end{eqnarray}
which is same as the state considered in \cite{Zeilinger}. The left particle flying in the IRSM $|S^\prime>>$ hits the screen in the region of interaction,
\begin{eqnarray}
 <<S^\prime|S^\prime>> = && <L_1|L_1> <R_1|R_1> + <L_2|L_2> <R_2|R_2> \nonumber\\
&+& <L_1|L_2> <R_1|R_2> + <L_2|L_1> <R_2|R_1>
\end{eqnarray}
So, it's clear that when the right particle is identified through which slit it has passed, i.e., $<R_1|R_2> = 0 $, then one sees clumps patterns formed by the left particles on the screen. On the other hand, if $<R_1|R_2> \ne 0 $, then interference occurs on the screen.

\begin{figure}[htbp]
\centerline{\includegraphics[width=0.8\textwidth]{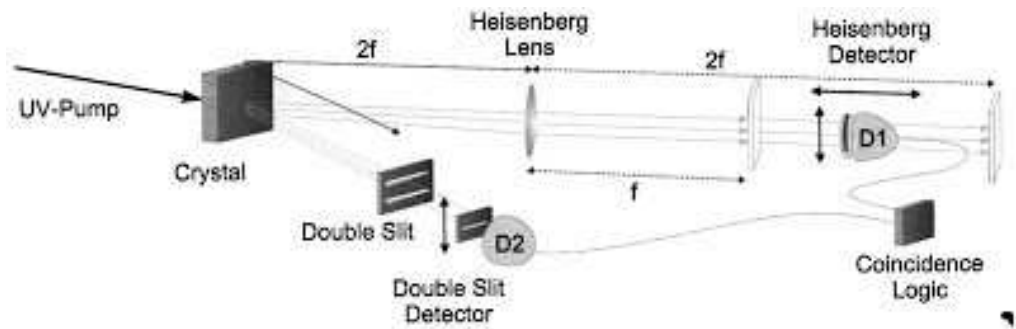}}
\caption{\textbf{Diagram of Dopfer's experiment}:~A pair of
momentum-entangled photons is created by the crystal and photon $1$
is observed by the detector $D1$ arranged behind the Heisenberg lens
while photon $2$ passes through a double-slit assembly and then
detected by $D2$. In the coincidence detection of photons $1$ and
$2$, an interference pattern or clump pattern is observed depending
on whether $D1$ is placed in the focal plane of the lens or in the
image plane.} \label{Figure4}
\end{figure}

In Dofper's experiment,  (Fig.~\ref{Figure4}), the moment a pair of momentum-entangled particles are created at the source, an IRSM, $|S>>$, gets excited in the entire apparatus. It  passes through the Heisenberg lens to the detector $D1$ and also passes through a double-slit assembly to $D2$. When the $D1$ is present at the focal plane of the lens, the components  passing through upper and lower slits overlap and the coincidence detection exhibits an interference. When $D1$ is placed at the image plane, then those components no more overlap and a clump pattern results.

Let us suppose that the separation between the source and double-slit is less than the distance between lens and source, i.e., $D2$ detects photon-2 much before the photon-1 passes through the lens. Now, place $D1$ at the focal plane and immediately after registering the photon at $D2$, quickly move it to the image plane and then register the photon-1. Repeat the same detection procedure for each and every entangled-pair generated at the source. Since $D1$ was at the focal plane when photon was registered at $D2$, an interference pattern will occur in the coincidence detection because the image of the double-slit interference forms on the image plane.

One can reverse this detection procedure i.e., initially place $D1$ at the image plane and record the photon at $D2$. Now, quickly move $D1$ to the focal plane and register photon-1. Since $D1$ was at the image plane during the detection of photon-2, the coincidence detection will not show the usual interference.

\section{Delayed choice quantum erasure experiment}
\begin{figure}[htbp]
\centerline{\includegraphics[width=0.6\textwidth]{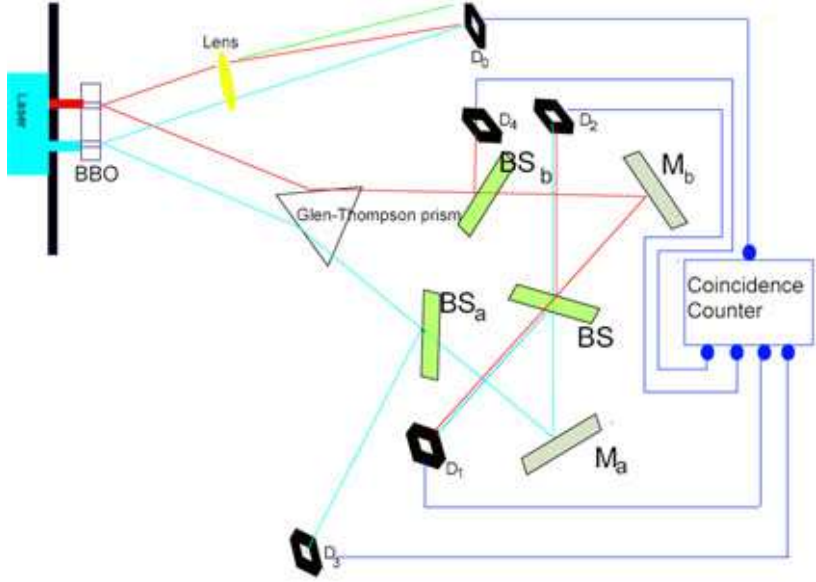}}
\caption{\textbf{Diagram of delayed choice quantum erasure
experiment}:~A laser pump shoots photons at double-slit assembly.
Immediately behind the slits, a beta barium borate crystal is placed
which can convert a single photon into two identical, orthogonally
polarized entangled photons with twice the wave length of the
original photon by a process called spontaneous parametric down
conversion. Much before a photon passes through the double-slit, its
IRSM induces two entangled states $|S_a> |S_b>$ and $|S_c> |S_d>$
from the upper and lower slit, respectively. One photon from an
entangled-pair reaches the movable detector $D_0$ while the other is
directed towards the Glan-Thompson prism. The photons reaching $D_0$
are called signal photons and those reaching $D_1$, $D_2$, $D_3$ and
$D_4$, via the 50:50 beam splitters $BS_a$, $BS_b$ and $BS$ and
100$\%$ reflecting mirrors $M_a$ and $M_b$, are called idler
photons. The optical path lengths are such that there is a $8$ns
time delay between first detecting a signal photon and later, its
entangled idler photon.} \label{Figure5}
\end{figure}

Consider the experimental setup as shown in Fig.~\ref{Figure5} and see reference \cite{Kim} for complete details.

The moment a photon is generated at the laser source, an IRSM gets excited in the entire space and also through the double-slit. Much before the photon passes through the double-slit assembly, the IRSM itself induces the process of spontaneous parametric down conversion in beta barium borate crystal and hence two secondary entangled modes, $|U>> (= |S_a> |S_b>$ and
$ |L>> (= |S_c> |S_d>)$, appear instantaneously in the entire apparatus; where, $|U>>$ and $|L>>$ are from upper and lower slits, respectively and obviously they are superposed. Suppose that the photon from source actually goes through the upper slit, then it gets split into two entangled photons such that one will be present in $|S_a>$ and the other one in $|S_b>$, respectively. However, it should be noted that even though photons are present in $|S_a>|S_b>$, the empty mode $|S_c>|S_d>$ is ontologically present and contributes to the observed phenomenon in the apparatus.

The superposed entangled IRSM, $|S_0>>$, is given by
\begin{eqnarray}
|S_0>> = |U>> + |L>> = |S_a> |S_b> + |S_c> |S_d>
\end{eqnarray}
Signal and idler photons simply fly in this IRSM and reaches the appropriate detectors through already predetermined entangled modes. So, it doesn't matter whether the signal photon is detected at $D_0$ before or after the detection of its entangled idler photon.

The modes $|S_b>$ and $|S_d>$ are further decomposed as
\begin{eqnarray}
|S_b> \rightarrow |S_b;4> + |S_b;1> + |S_b;2> \quad ;\,\,|S_d> \rightarrow |S_d;3> + |S_d;1> + |S_d;2> \nonumber
\end{eqnarray}
where, $|S_b;4>$ is the component of the mode $|S_b>$ reaching the detector 4, etc.,
and since they are entangled like in Eq. (\ref{ESDC}), one has
\begin{eqnarray}
|S_a>|S_b> = |S_a;4>|S_b;4> + |S_a;1>|S_b;1> + |S_a;2>|S_b;2>
\end{eqnarray}
and
\begin{eqnarray}
|S_c>|S_d> = |S_c;3>|S_d;3> + |S_c;1>|S_d;1> + |S_c;2>|S_d;2>
\end{eqnarray}

Therefore, the observed photon distribution at $D_0$ can be written in terms of the distributions at $D_1$, $D_2$, $D_3$ and $D_4$ as
\begin{eqnarray}
<<S_0|S_0>>  &= &<S_a;4|S_a;4> <S_b;4|S_b;4> + <S_c;3|S_c;3> <S_d;3|S_d;3> \nonumber\\
 &+& <S_a;1|S_a;1> <S_b;1|S_b;1> + <S_c;1|S_c;1> <S_d;1|S_d;1> \nonumber\\
&+& <S_c;1|S_a;1> <S_d;1|S_b;1> + <S_a;1|S_c;1> <S_b;1|S_d;1> \nonumber\\
&+& <S_a;2|S_a;2> <S_b;2|S_b;2> + <S_c;2|S_c;2> <S_d;2|S_d;2> \nonumber\\
&+& <S_c;2|S_a;2> <S_d;2|S_b;2> + <S_a;2|S_c;2> <S_b;2|S_d;2>
\end{eqnarray}

It's clear from the above expression that signal photons whose entangled idler was detected at $D_3$ or $D_4$ show single slit diffraction whereas the idlers detected at $D_1$ or $D_2$ exhibit interference. Everything is causal and there is nothing like retrocausal influences.

\section{Delayed Choice Entanglement Swapping}
Two well-separated photons can be made to become entangled even though they have never interacted and shared any common past and is known as entanglement swapping. To achieve this, two pairs of entangled photons are produced and one photon from each pair is sent to Alice and Bob, respectively. The remaining two photons are sent to Victor. Now, Alice's and Bob's photons can be made to become entangled by simply projecting Victor's photons onto an entangled state \cite{EPR, Schro, Zuko}. Here, an interesting aspect is that this entanglement swapping can be delayed \cite{Peres,Cohen,Ma}. First, Alice and Bob measure polarization states of their photons in their own respective basis. Later, Victor randomly chooses a basis which is either entangled or uncorrelated, to project his photons. Even though Victor has recorded the polarization states of his photons much later than Alice and Bob, but still his future choice of basis seems to dictate whether Alice's and Bob's photons to become entangled or not. See Ref. \cite{Ma} for more explanation and diagrammatic details.

Our common sense experience is that the `present' is simply a resultant outcome of already happened and no more existing `past'. Therefore, any theory representing or reflecting the reality of Nature must be casual. Even though the underlying equations describing the physical process may have time-reversal invariance, but its solutions should not be. Exactly in this sense, quantum mechanics is reflecting the casual reality of Nature at more fundamental quantum level.

According to the present non-dualistic picture, it is the measurement of Alice and Bob casually determines the outcome of Victor but not vice versa. This is shown by making use of the mechanisms from both the Wheeler's delayed choice experiment and the spooky action at a distance.

Let $|\phi_{12}>>$ and $|\phi_{34}>>$ be the two entangled states of two pairs of photons. Photon 1 in $|\phi_{12}>>$ is send to Alice and photon 4 in $|\phi_{34}>>$ is sent to Bob. Photons 2 and 3 are send to Victor.

Let $|P^{A +}_1> $ and  $|P^{A -}_1>$ be Alice's basis for measuring photon 1 and  $|P^{B +}_4>$ and $|P^{B -}_4>$ be Bob's basis for measuring photon 4. Then the entangled states
$|\phi_{12}>>$ and $|\phi_{34}>>$ can be expressed in Alice and Bob basis respectively as
\begin{eqnarray}
|\phi_{12}>> = <P^{A +}_1|\phi_{12}>> |P^{A +}_1> + <P^{A -}_1|\phi_{12}>> |P^{A -}_1>
\end{eqnarray}
and
\begin{eqnarray}
|\phi_{34}>> = <P^{B +}_4|\phi_{34}>> |P^{B +}_4> + <P^{B -}_4|\phi_{34}>> |P^{B -}_4>
\end{eqnarray}
If Alice observes the photon 1 in the state $|P^{A +}_1> $, then the photon 2 is thrown into the state $<P^{A +}_1|\phi_{12}>>$ due to spooky action at a distance. Alice could have as well observed the photon 1 in $|P^{A -}_1>$. So, the photon 2 state depends on Alice's observation on photon 1 and similarly, photon 3 state on Bob's measurement.
The state of photon 2 after Alice's measurement can be written as
\begin{eqnarray}
<P^{A \pm}_1|\phi_{12}>> = |P^{A_1 \pm}_2>
\end{eqnarray}
where the state $|P^{A_1 +}_2>$ can be read as `the state acquired by photon 2 when Alice observes photon 1 in the state $|P^{A +}_1>$. Similarly, the state of photon 3 after Bob's measurement can be written as
\begin{eqnarray}
<P^{B \pm}_4|\phi_{34}>> = |P^{B_4 \pm}_3>
\end{eqnarray}

Therefore, the joint state of photon 2 and 3, which Victor encounters, is given by
\begin{eqnarray}
|P^{A_1 \pm}_2> |P^{B_4 \pm}_3> \equiv |P^{A_1 \pm}_2;P^{B_4 \pm}_3>>
\end{eqnarray}

Let $|V_{23}^+>> $ and $|V_{23}^->> $ be the basis in which Victor makes a joint measurement on photon 2 and 3. Then, one has
\begin{eqnarray}
|P^{A_1 \pm}_2;P^{B_4 \pm}_3>> =  <<V_{23}^+|P^{A_1 \pm}_2;P^{B_4 \pm}_3>> |V_{23}^+>>  + <<V_{23}^-|P^{A_1 \pm}_2;P^{B_4 \pm}_3>> |V_{23}^->> \nonumber
\end{eqnarray}

Hence, the frequency of detection for Victor is
\begin{eqnarray}
C^{\pm}_{V_{23}}(A_1 \pm;B_4 \pm) = |<<V_{23}^{\pm}|P^{A_1 \pm}_2;P^{B_4 \pm}_3>>|^2 \label{dce}
\end{eqnarray}
which clearly depends on already recorded measurements of Alice and Bob on photon 1 and 4, respectively. Further, Victor can change his basis randomly like in the case of Wheeler's delayed choice experiment. One can easily check that the Eq. (\ref{dce}) is in exact agreement with the experimental results found in Ref. \cite{Ma}.

In the same manner, the delayed choice entanglement swapping in time \cite{DCEST} can also be explained. First, an entangled pair of photons, in the state $|\phi_{12}>>$, is created and photon 1 is sent to Alice who makes a measurement in her basis.
\begin{eqnarray}
|\phi_{12}>> = <P_1^{A +}|\phi_{12}>> |P_1^{ A +}> + <P_1^{A -}|\phi_{12}>> |P_1^{ A -}>
\end{eqnarray}
As a consequence of spooky action at a distance, photon 2 is thrown into a definite state depending on the out come of Alice measurement, given by
\begin{eqnarray}
|P^{A_1 \pm}_2> = <P_1^{A \pm}|\phi_{12}>>
\end{eqnarray}
At a later time, another entangles pair of photons in the state $|\phi_{34}>>$ is created and photon 3 along with the photon 2 is send to Victor for joint measurement. The joint state encountered by Victor is
\begin{eqnarray}
|P_2^{A_1 \pm}> |\phi_{34}>> \equiv |P_2^{A_1 \pm}; \phi_{34}>>>
\end{eqnarray}
Now, Victor makes a joint measurement on photon 2 and 3 in his basis,
\begin{eqnarray}
|P_2^{{A_1} \pm}; \phi_{34}>>> =  <<V^+_{23}|P_2^{A_1 \pm}; \phi_{34}>>> |V^+_{23}>>  + <<V^-_{23}|P_2^{A_1 \pm}; \phi_{34}>>> |V_{23}^->>
\end{eqnarray}
and finds the photon 2 and 3 either in one of the basis which determines the state of the photon 4,
\begin{eqnarray}
|P_4^{{V_{23}} \pm};A_1 \pm> =  <<V^{\pm}_{23}|P_2^{A_1 \pm}; \phi_{34}>>>
\end{eqnarray}
and Bob finally measures this state in his basis
\begin{eqnarray}
|P_4^{V_{23} \pm};A_1 \pm> = <P_4^{B +} |P_4^{V_{23} \pm};A_1 \pm> |P_4^{B +}>  +   <P_4^{B -} |P_4^{V_{23} \pm};A_1 \pm> |P_4^{B -}>
\end{eqnarray}
and calculates the frequency of detection at the end of the experiment as,
\begin{eqnarray}
C^{B_4 \pm}_4(A_1 \pm: V_{23} \pm) = |<P_4^{B \pm} |P_4^{V_{23} \pm};A_1 \pm>|^2 \label{dct}
\end{eqnarray}
which clearly depends on in which state Alice has seen the photon 1 and Victor has seen the photons 2 and 3. ote that, both Victor and Bob are free to chose their basis randomly and Eq. (\ref{dct}) is in exact agreement with the results found in Ref. \cite{DCEST}.

\section{The Hanbury-Brown-Twiss effect}
The Hanbury-Brown Twiss effect describes the interference of incoherent radiation \cite{HBT1,HBT2}. Two sources $a$ and $b$, separated by some distance, emits particle of same energy
and they fly in their IRSMs $|a>$ and $|b>$, respectively. Away from these sources, two detectors $A$ and $B$ make a two-particle coincidence measurement. If $\hat{H}_a |a> = E |a>$ and $\hat{H}_b |b> = E |b>$; where $\hat{H}_a$ and $\hat{H}_b$ are Hamiltonians of particle from $a$ and $b$ with energy eigenvalue $E$, then one needs two particle eigenstate of eigenvalue $2E$ for joint-detection to happen. Such a state obeys the following eigenvalue equation
\begin{eqnarray}
({\hat{H}}_a + \hat{H}_b ) |a>|b> = 2E |a>|b>
\end{eqnarray}
Let $\hat{P}_{A} = |A><A|$ and $\hat{P}_{B} = |B><B|$ be the operators associated with $A$ and $B$. Then the joint detection is described by the operator
\begin{eqnarray}
\hat{P}_{AB} = |A><A| \otimes |B><B|
\end{eqnarray}
Keeping in mind that the detectors are insensitive to a particular particle from particular source because they see only energy, one can write down
\begin{eqnarray}
|a> |b> = |\phi_+>> + |\phi_->>
\end{eqnarray}
where,
\begin{eqnarray}
|\phi_{\pm}>> = (|a> |b> \pm |b>|a>)/2
\end{eqnarray}
Here $|\phi_+>>$ and $|\phi_->>$ corresponds to symmetric and anti-symmetric states. Since they are mutually orthogonal, both particles must be present either in $|\phi_+>>$ or $|\phi_->>$ for joint detection. Therefore, the mode which contributes at $AB$ is
\begin{eqnarray}
|{\tilde{\phi}}_{\pm}>> = \hat{P}_{AB} |\phi_{\pm}> = (<A|a> <B|b> |A>|B> \pm <B|a> <A|b> |B>|A>)/2
\end{eqnarray}
and its interaction with the excited dual in the detector as
\begin{eqnarray}
<<{\tilde{\phi}}_{\pm}|{\tilde{\phi}}_{\pm}>> &= & |<A|a>|^2 |<B|b>|^2 \pm <b|A> <a|B> <A|a> <B|b>  \nonumber \\
&+& |<B|a>|^2 |<A|b>|^2 \pm <a|A><b|B> <A|b><B|a>
\end{eqnarray}
where, one can take $<A|A> = <B|B> = <A|B> = <B|A> = 1$ if both detectors are identical. The above is the frequency of joint detection. Note that, both IRSMs $|a>$ and $|b>$ influences both detectors $A$ and $B$ simultaneously. Even if the particle in $|a>$ is detected by $A$, its influence is felt by the detector $B$ and vice versa.

\section{Non-linear Schr\"odinfer equation}
Consider again the inner product, $<S|S>$, between the dual, $<S|$, in the detector space and the original IRSM, $|S>$. This corresponds to just an ideal situation. But, in geral the inner-product can be equal to $\pm \eta <S|S>$; where, $\eta$ can be related to the detection efficiency and its sign to detection convention. Let us assume a `free particle' in one dimension along $x$-axis inside a medium which continuously interact with the IRSM. Then,
\begin{eqnarray}
<S|S> = \eta \int dx |<x|S>|^2 = \eta \int dx |\psi(x)|^2
\end{eqnarray}
and this interaction at a given point, $x$, can be treated as a potential in the Schr\"odinger equation,
\begin{eqnarray}
H = \frac{{\hat{p}^2}}{2 m} + \eta |\psi(\hat{x})|^2
\end{eqnarray}
Since, the Hamiltonian is Hermitian, one obtains the time-dependent non-linear Schr\"odinger wave equation \cite{NonSch} as
\begin{eqnarray}
- \frac{\hbar^2}{2 m} \frac {\partial^2 \psi} {\partial {x^2}} + \eta |\psi|^2 \psi = i \hbar \frac{{\partial} \psi}{\partial t}
\end{eqnarray}

Before concluding the present paper, it's important to quote Einstein, ``Reality is out there independent of our observation''. Infact any reasonable theory must aim at such a description of Nature, which anyway will yield the correct results upon observation. The quantum mechanical formalism works exactly in this manner as shown by the present non-dualistic interpretation.

It's wrong to explain any quantum phenomenon by invoking Heisenberg's uncertainty principle, though those bounds imposed by it does exist in Nature. Such an explanation is sure to blur our vision to capture the actual picture reality as anticipated by Einstein. Further, the uncertainty principle is a consequence of the quantum formalism but not vice versa. Application of uncertainty relation to explain quantum phenomena results in weird interpretation of reality like, `Nature is inherently uncertain', `probability is an intrinsic property of Nature', `even Nature doesn't know through which slit a particle actually goes through'...etc. If Nature does not know and the particle also does not know (because particle is a part of Nature), then how probabilities and uncertainties arise in the first place is questionable.

In the present non-dualistic picture, a free particle moves in its IRSM of a definite wavelength which can not be confined inside a width, $\Delta x$, less than its wave length. Any such attempt will change the IRSM itself. I present a naive mathematical argument as follows;  we have a minimum width $\Delta {x_{min}}$ inside which a particle can be confined, without any change in its momentum eigenvalue, is $\Delta {x_{min}} = \lambda $ which yields $ p \,\, \Delta {x_{min}} = h$, by using the de Brogle's relation. But, inside that width, the point particle still moves with the same momentum..

If the IRSM encounters a vertical slit of width $\Delta y$, then from Bragg's law, for first order diffraction,
$$ \Delta y \,\, \sin{\theta} = \lambda  \Rightarrow \Delta y \,\, p\,\,\sin{\theta} = h \Rightarrow \Delta y \,\, p\,\, \geq h$$
So, there exist a maximum momentum, $P_{max} = h/{\Delta y}$, a particle can have such that one can observe at least first order diffraction. Any momentum greater than $P_{max}$ will not exhibit any diffraction except the zeroth order. Therefore, confining and observing a particle are not the same thing. When a particle is confined, then it will be still moving within its IRSM. But, when it is detected means that it appeared at some location in the region of $<S|S>$. So, Heisenberg's uncertainty relations have a different kind of meaning in the present non-dualistic interpretation.

In conclusion, I have interpreted the Schr\"odinger wave function or equivalently the state vector of a quantum particle as an Instantaneous Resonant Spatial Mode (IRSM) and shown how the quantum formalism is related to the Nature of reality. Both particle and its IRSM are created simultaneously such that the particle flies in IRSM. This non-dualistic picture is analogues to the objects moving in the curved space-time of general theory of relativity. Thus, wave function does not have any resemblance to classically known waves though it obeys   Schr\"odinger's wave equation. Unlike in Newtonian mechanics, a complete knowledge about the initial state and hence the state at any later moment is unavailable to experimental observation though Nature Herself is aware of it. This is because of the fact that the particles are not living in $R^3$ Euclidean space and also due to the inner-product interaction of IRSM. Most importantly, quantum mechanics itself is not a probabilistic theory since all the quantum phenomenon considered in the present paper were explained at a single quantum mechanical level. It was shown that the Born's probabilities are equivalent to observed frequencies arising due to the nature of doing the experiments. Einstein was indeed correct in saying, " God does not play dies".

The most important conclusion one can draw from non-duality is that Nature is not retrocasual and respects casuality at least at the level of quantum mechanics. Towards the end, I remarked about the HBT effect and also pointed out on the possible origin of non-linear Schr\"odinger equation. The explanations given for the Young's double-slit experiment, Wheeler's delayed  choice experiment and spooky action at a distance at a single quantum level seems to be sufficient for the unambiguous understanding of all known quantum mechanical phenomena so far.

Since the present non-dualistic interpretation is a visualization of nature of reality reflected within the quantum formalism, it will go through both time-dependent and relativistic quantum mechanics. In the relativistic case, the IRSM is such that, apart from obeying the usual quantum mechanical commutation relations, it takes care of the cosmic speed limit for its resonant particle, though it itself can change instantaneously. Without much difficulty, it can be seen that the registered physical phenomena are independent of relative frame of reference. Though at the present moment I reserve my comments on taking this wave-particle non-duality over to quantum field theories, nevertheless, it should naturally go through and may shed some new light, particularly in the context of renormalization procedure and hence to understand the quantum nature of Einstein's gravity. Finally, the important essence of the present paper is, "Nature does not have to play dies in order to run our casual quantum mechanical Universe".



\begin{references}

\bibitem{Feyn1} R. P. Feynmann, The Feynman Lectures on Physics,  Volume 3, Addison‐Wesley, 1963.

\bibitem{Feyn2} R. P. Feynmann, Simulation physics with Computers, Journ. Th. Phys. 21, 467 (1982).

\bibitem{spi1} C.  Josson, Electron Diffraction at Multiple Slits, Am. J. Phys, 42, 4 (1974).

\bibitem{spi2} A. Zeilinger, R. Gahler, C.G. Shull,  W. Treimer and W. Mampe, Single- and double-slit diffraction of neutrons, Rev. Mod. Phys. 60, 1067 (1988).

\bibitem{spi3} O. Carnel and J. Mlynek, Young’s double-slit experiment with atoms: A simple atom interferometer, Phys. Rev. Lett. 66, 2689 (1991).

\bibitem{spi4} W. Schollkopf and J.P. Toennies, Nondestructive Mass Selection of Small van der Waals Clusters, Science, 266, 1345 (1994).

\bibitem{spi6} O. Nairz, M. Arndt and A. Zeilinger, Quantum interference experiments with large molecules, Am.  J.  Phys. 71, 319 (2003).

\bibitem{Wheeler} J.A. Wheeler and W.H. Zurek, Quantum Theory and Measurement, 182-213, (Princeton University Press, 1984).

\bibitem{Born} M. Born, The statistical interpretation of quantum mechanics - Nobel Lecture, December 11, {1954}.

\bibitem{GRW} G.C. Ghirardi, A. Rimini and T. Weber, Unified dynamics for microscopic and macroscopic systems, Phys. Rev. D. 34, 470 (1986).

\bibitem{Au} G. Auletta, Foundations and Interpretation of Quantum Mechanics, (World Scientific, 2001).

\bibitem{Bohm} D. Bohm, A Suggested Interpretation of the Quantum Theory in Terms of "Hidden" Variables, Phys. Rev. 85, 166-193 (1952).

\bibitem{SAD1} S.J. Freedman and J.F. Clauser, Experimental test of local hidden-variable theories, Phys. Rev.  Lett. 28, 938 (1972).

\bibitem{SAD2} A. Aspect, J. Dalibard and G. Roger, Experimental test of Bell’s inequalities using time-varying analyzers, Phys. Rev. Lett. 49, 1804 (1982).

\bibitem{SAD3} G. Weihs, T. Jennewein,  C. Simon, H. Weinfurter and A. Zeilinger, Violation of Bell’s inequality under strict Einstein locality conditions, Phys. Rev. Lett. {81},
5039 (1998).

\bibitem{SAD4} M. Rowe et al., Experimental violation of a Bell’s inequality with efficient detection, Nature. 409, 791 (2001).

\bibitem{SAD5} M. Ansmann et al., Violation of Bell’s inequality in Josephson phase qubits, Nature. 461, 504 (2009).

\bibitem{SAD6} J. Hofmann et al., Heralded entanglement between widely separated atoms, Science. 337, 72 (2003).

\bibitem{SAD7} M. Giustina et al., Bell violation using entangled photons without the fair-sampling assumption, Nature. 497, 227 (2013).

\bibitem{SAD8} B. Hensen et al., Loophole-free Bell inequality violation using electron spins separated by 1.3 kilo-metres, Nature. 526, 682 (2015).

    \bibitem{Dirac} P.A.M. Dirac, The Principles of Quantum Mechanics, 16 (Oxford University Press, 1958).

\bibitem{Sakurai} J.J. Sakurai, Modern Quantum Mechanics, 33-34, (Addison‐Wesley, 1994).

\bibitem{EPR} A. Einstein, B. Podolsky and N. Rosen, Can Quantum-Mechanical Description of Physical Reality Be Considered Complete?, Phys. Rev. 47, 777 (1935).

\bibitem{Bell} J.S. Bell, On the Einstein Podolsky Rosen Paradox, Physics. 1, 195 (1964).

\bibitem{Jacques} V. Acques et al., Experimental realization of Wheeler’s delayed-choice GedankenExperiment, Science. 315, 966 (2007).

\bibitem{SingleAtom} A.G. Manning, R.I. Khakimov, G. Dall and A.G. Truscott, Wheeler’s delayed-choice gedanken experiment with a single atom, Nature Physics. 11, 539 (2015).

\bibitem{Zeilinger} A. Zeilinger, Experiment and the foundations of quantum physics, Rev. Mod. Phys. 71, 5288 (1999).

\bibitem{Kim} Y-H. Kim et al., A Delayed Choice Quantum Eraser, Phys. Rev. Lett. 84, 1 (2000).

\bibitem{Schro} E. Schr\"{o}dinger, Quantum Theory and Measurement, 152-167, (Princeton University Press, 1984).

\bibitem{Zuko} M. Zukowski, A. Zeilinger, M.A. Horne and A.K. Ekert, ‘‘Event-ready-detectors’’ Bell experiment via entanglement swapping, Phys. Rev. Lett. 71,  4287 (1993).

\bibitem{Peres} A. Peres, Delayed choice for entanglement swapping, J. Mod. Opt. 47, 139 (2000).

\bibitem{Cohen} O. Cohen, Counterfactual entanglement and nonlocal correlations in separable states, Phys. Rev. A. 60, 80 (1999).

\bibitem{Ma} X-S. Ma, Experimental delayed-choice entanglement swapping, Nature. Phys. 8, 480, (012).

\bibitem{DCEST} E. Magadish et al., Entanglement Swapping between Photons that have Never Coexisted, Phys. Rev. Lett. 110, 210403 (2013).

\bibitem{HBT1} R. Hanbury Brown and R.Q. Twiss, Correlation between Photons in two Coherent Beams of Light, Nature. 177, 27 (1956).

\bibitem{HBT2} R. Hanbury Brown and R.Q. Twiss, The Question of Correlation between Photons in two Coherent Beams of Light, Nature. 178, 1447 (1956).

\bibitem{NonSch} B. Malomed, Encyclopedia of Nonlinear Science, 639-643 (Scott, Alwyn, 2005).


\end{references}
\end{document}